\documentclass[12pt,a4,epsf]{article} \usepackage{cite}
\usepackage{graphicx} \usepackage{amssymb} \global\arraycolsep=2pt
\input{epsf} 
\usepackage{graphicx}
\def\@fmsl@sh#1#2#3{\m@th\ooalign{$\hfil#1\mkern#2/\hfil$\crcr$#1#3$}}
 \def\eq#1\en{\begin{equation}#1\end{equation}}
\def\s[#1,#2]{[#1\stackrel{\star}{,}#2]}
\def\sx[#1,#2]{[#1\stackrel{\star_{x}}{,}#2]} 
\begin{document}
\makeatletter
\def\fmslash{\@ifnextchar[{\fmsl@sh}{\fmsl@sh[0mu]}}
\def\fmsl@sh[#1]#2{%
  \mathchoice
    {\@fmsl@sh\displaystyle{#1}{#2}}%
    {\@fmsl@sh\textstyle{#1}{#2}}%
    {\@fmsl@sh\scriptstyle{#1}{#2}}%
    {\@fmsl@sh\scriptscriptstyle{#1}{#2}}}
\def\@fmsl@sh#1#2#3{\m@th\ooalign{$\hfil#1\mkern#2/\hfil$\crcr$#1#3$}}
\makeatother

\thispagestyle{empty}
\begin{titlepage}
\begin{flushright}
CALT-68-2510\\
June, 28 2004
\end{flushright}

\vspace{0.3cm}
\boldmath
\begin{center}
  \Large {\bf Minimal Grand Unification Model \\ in an Anthropic Landscape}
\end{center}
\unboldmath
\vspace{0.8cm}
\begin{center}
  {\large Xavier Calmet\footnote{
email:calmet@theory.caltech.edu}}\\ \end{center}
\begin{center}
{\sl California Institute of Technology, Pasadena, 
California 91125, USA}
\end{center}
\vspace{\fill}
\begin{abstract}
\noindent
It has been recently pointed out by Arkani-Hamed and Dimopoulos that
if the universe is a landscape of vacua, and if therefore fine-tuning
is not a valid guidance principle for searching for physics beyond the
standard model, supersymmetric unification only requires the fermionic
superpartners. We argue that in that landscape scenario, the fermionic
superpartners are not needed for unification, which can be achieved
in SO(10) either via a direct breaking to the standard model at the
grand unification scale or through an intermediate gauge symmetry. In
most minimal SO(10) models, the proton lifetime is long enough to
avoid the experimental bounds. These models are the truly minimal
fine-tuned extensions of the standard model in the sense proposed by
Davoudiasl et al..
\end{abstract}  
\end{titlepage}
\section{Introduction}

It has been pointed out a few years ago that the fine-tuning problem
of the standard model, i.e. why the Higgs boson's mass is so small,
and thus stable with respect to radiative corrections, in comparison
to the Planck scale (this problem is also referred to as the
naturalness problem \cite{'tHooft:1980xbis}), could be explained by
the anthropic principle \cite{Agrawal:1997gf}. The non-vanishing value
of the cosmological constant can be see as a failure of the
fine-tuning problem as a guidance for physics beyond the standard
model.  Indeed if for example a symmetry, e.g. supersymmetry, was to
explain the magnitude of the cosmological constant it would require a
breakdown of the physical theories we are so familiar with at a scale
of the order of $10^{-3}$ eV and new physics would probably have been
observed already.

This could imply that fine-tuning is not a valid physical question and
that indeed as in any renormalizable theory the gauge and Yukawa
couplings are parameters that have to be measured and whose magnitudes
large or small cannot be explained from first principles. In that case
it makes little sense to discuss whether a certain value of a given
parameter is natural or not.  The remaining problem is then to
understand the splitting between the Planck or grand unified scale and
the weak scale, the so-called gauge hierarchy problem. This is the
approach that has been advertized in \cite{Calmet:2002rf} where it was
shown that a seesaw mechanism in the Higgs sector of a simple
extension of the standard model can explain the magnitude of the
electroweak scale and also trigger the Higgs mechanism. Accidental
cancellations of radiative corrections involving large fine-tuning are
also conceivable \cite{Calmet:2003uj}. The discovery at the LHC of a
single Higgs boson and the lack of any new physics signal would
confirm the cosmological constant hint that the fine-tuning issue is
irrelevant. We nevertheless note that a fat Higgs model
\cite{Harnik:2003rs} or a composite Higgs model see
e.g. \cite{Calmet:2000th,Hong:2004td} could be a valid alternative to
the standard model if only one Higgs boson was discovered at the LHC.

On the other hand one could argue that the fine-tuning issue is not
irrelevant, but was badly formulated. It has been recently discovered
that string/M theory has a landscape of vacua
\cite{Douglas:2003um}. At first sight this sounds like a disaster for
the leading candidate for a theory of everything, but if one accepts
the anthropic principle as a valid scientific explanation this
abundance of vacua allows to rephrase the fine-tuning
problem. Basically the question becomes: given the probability
distribution of the physical parameters of a theory, what is the
probability that we happen to live in a universe, or vacuum, that has
a given set of parameters and how probable is that vacuum? This
question makes sense if indeed no physical principle is found in
string/M theory to select a particular vacuum and if indeed string/M
theory is the correct theory of nature. It should be noted that a
cosmological constant of the right order of magnitude had been
predicted by Weinberg \cite{Weinberg:dv} already a long time ago using
anthropic considerations. Recently such considerations have been
applied to the supersymmetry breaking scale \cite{Susskind:2004uv}.

This reformulation of the fine-tuning problem was the motivation for
Arkani-Hamed and Dimopoulos \cite{Arkani-Hamed:2004fb} to propose that
supersymmetry might not be required to explain a light Higgs boson but
can be useful to explain the unification of the gauge couplings, see
also \cite{Arvanitaki:2004eu}. The basic idea is to break
supersymmetry at a very high scale, thereby giving large masses to all
the scalar fields with the exception of the standard model Higgs boson
whose mass is assumed to remain light and thus has to be
fine-tuned. The new fermions appearing in the supersymmetric extension
of the standard model are assumed to remain light enough so that they
can contribute, as usually assumed, to the running of the gauge
couplings and guaranty the unification of the couplings. It should be
emphasized that in this framework, unification does not require an
additional fine-tuning since the fermion masses could be protected by
chiral symmetries. While this scenario is interesting from the
phenomenological perspective (it predicts a plethora of new phenomena
at the LHC,) it is clearly not a minimal extension of the standard
model that leads to the unification of the gauge couplings.

Motivated by this study and the minimal extension of the standard
model proposed by Davoudiasl et al. \cite{Davoudiasl:2004be}, we
consider the minimal grand unified theory in an anthropic landscape
scenario. We point out that supersymmetry, or in other word the new
fermions present in that scenario, is not required for grand
unification. In the remaining of this work, we will describe the truly
minimal grand unified model that accounts for all the, to date,
observed phenomena. The simplest viable model is based on a SO(10)
group \cite{Fritzsch:nn}. We will assume that fine-tuning is a not valid
physical question or rather that it should be rephrased in terms of the
landscape problematic. One should nevertheless keep in mind that the
anthropic principle could apply only to certain parameters of the
theory, e.g. the cosmological constant. It the sequel we will ignore
this possibility and assume that fine-tuning is acceptable for all the
parameters of the model.

\section{Fine-tuned minimal SO(10) grand unification}

We will be looking for a unified model with the following properties:
\begin{itemize}
\item[a)] numerical unification of the gauge couplings,
\item[b)] enough free parameters to fit the fermion masses,
\item[c)] baryogenesis,
\item[d)] long lived proton to avoid the experimental bounds.
\end{itemize}
The dark matter, the inflaton, the cosmological constant and gravity
are considered to be different sectors in the spirit of
ref. \cite{Davoudiasl:2004be} and we shall not try to unify these with
the remaining interactions. These sectors of the theory are assumed to be
described by the minimal models presented in
\cite{Davoudiasl:2004be}. We will argue that non-supersymmetric SO(10)
grand unified models are viable candidates for grand unification. It
should be nevertheless mentioned, that on the contrary to
supersymmetric theory, there is no obvious dark matter candidate in
these models. Obviously one can easily either introduce a singlet
under SO(10) to describe dark matter or introduce some scalar
multiplet in a representation of SO(10) that does not spoil the
unification of the standard model gauge couplings.

There are different ways to break the grand unified group SO(10) to
the standard model group SU(3)$\times$SU(2)$\times$U(1), we shall
present four different non-supersymmetric models that are
phenomenologically viable in the sense that the proton lifetime is
long enough to escape discovery and that the gauge couplings of the
standard model unify.

Baryogenesis in non-supersymmetric SO(10) grand unified theories can
happen at the grand unification scale \cite{Abbott:1982hn} as one of
the predictions of the model is that the baryon number is
violated. But, a potentially serious problem with that scenario is the
low value of the thermalization temperature of the inflaton. The
unification scale is expected to be of the order of $10^{15}$
GeV. Heavy Higgs bosons and gauge bosons leading to baryon decay are
expected to have a mass of the order of the grand unification scale to
avoid problems with the bounds on proton decay. They thus have a mass
greater than that of the inflaton and it seems kinetically impossible
to produce them directly through inflaton decay. However one can
imagine a scenario where grand unified gauge and Higgs bosons leading
to baryon decay are produced non-thermally \cite{Kolb:1996jt}. We
shall nevertheless invoke leptogenesis \cite{Fukugita:1986hr} as our
mechanism to generate the baryon asymmetry as it seems more plausible
and appears automatically in the minimal SO(10) model.

As mentioned previously, leptogenesis is not a requirement, but one
has to go beyond the {\bf 10} to get a realistic spectrum for the
fermion masses. One can introduce for example the {\bf 126}. If $D$
parity (a $Z_2$ discrete symmetry contained in SO(10)) is broken at
the grand unified scale, the seesaw formula appears naturally. Note
that the seesaw mechanism is not imposed to explained the smallness of
the neutrino masses or leptogenesis, but it just follows from the
minimalistic assumption. The {\bf 126} generates Majorana masses for
the neutrinos and baryogenesis happens through leptogenesis.

\subsection{Minimal SO(10) grand unification model}

The first model is that proposed by Lavoura and Wolfenstein
\cite{Lavoura:su}. It is the minimal SO(10) grand unification model
broken directly at the grand unification scale to the standard model
gauge group. The fermion masses are generated by the Higgs multiplets
in the ${\bf 10}$, ${\bf 126}$ and ${\bf 210}$ representations.  The
Higgs bosons in the ${\bf 10}$ and ${\bf 126}$ representations break
SO(10) to SU(2)$_L \times$SU(2)$_R \times$SU(4)$_{PS}$. The (2,2,1) of
the ${\bf 10}$ gets a vacuum expectation value $v_1$, the (1,3,10) of
the ${\bf 126}$ gets a vacuum expectation value $v_R$, the (2,2,15) of
the ${\bf 126}$ gets a vacuum expectation value $v_{15}$ with
$v_{15}<v_R$ and the (1,1,1) of the ${\bf 210}$ gets a vacuum
expectation value $v_U$ with $v_U={\cal O}(M_U)$, where $M_U$ is the
grand unification scale. Obviously one has to impose $v_U>v_R$. These
are the requirements to fit the fermion masses and to unify the gauge
couplings at a scale $M_U$, see \cite{Lavoura:su} for details. 

Lavoura and Wolfenstein have shown that if the heavy higgs bosons and
heavy gauge bosons have masses smaller (e.g. a factor 30) than the
energy scale where the gauge symmetry is broken, then the running of
the gauge couplings can be significantly affected and grand
unification is possible for a range of parameters. The mechanism
proposed in \cite{Lavoura:su} is based on the observation that there
are nine gauge bosons in SO(10) that are not standard model gauge
bosons but that do not lead to proton decay which are assumed to have
masses of the order of $v_R=M_R$. The main idea is to use these gauge
bosons that do not lead to proton decay to affect the running of the
gauge couplings significantly, whereas the remaining non-standard model
gauge bosons of SO(10) that lead to proton decay are assumed to have a
mass of the order of the grand unification scale $M_U$.

To illustrate our point, we shall use the one loop results derived in
\cite{Lavoura:su}:
\begin{eqnarray}
\ln{\frac{M_U}{M_Z}}&=& \frac{1}{128} \left [ 60 \pi \omega_1(M_Z)-60
\pi \omega_2(M_Z) + 5 (\lambda_1 -\lambda_2) \right ]
\\
\omega_G(M_U)&=&\frac{1}{128} \left [ 95 \omega_1(M_Z)+123
\omega_2(M_Z) +  \frac{95 \lambda_1 +123 \lambda_2}{12 \pi} \right ]
\\
\omega_3(M_Z)&=&\frac{1}{128} \left [ -115 \omega_1(M_Z)+333
\omega_2(M_Z) +  \frac{-115 \lambda_1 +333 \lambda_2-218 \lambda_3}{12 \pi} \right ]
\end{eqnarray}
where $\omega_1$ is the inverse of the U(1) gauge coupling, $\omega_2$
is the inverse of the SU(2) gauge coupling, $\omega_3$ is the inverse
of the SU(3) gauge coupling and $\omega_{G}$ is the inverse of the
SO(10) gauge coupling. The $\lambda$'s represent the contributions of
the heavy gauge bosons and higgs bosons to the running of the gauge
couplings and are given by \cite{Lavoura:su}:
\begin{eqnarray}
\lambda_1&=& 8+\frac{294}{5} 
\ln{\frac{M_U}{M_R}} 
- \frac{274}{5} \ln{\frac{M_U}{M_1}}
- \frac{142}{5} \ln{\frac{M_U}{M_2}}
- \frac{36}{5} \ln{\frac{M_U}{M_3}}
\\ \nonumber &&
- \frac{114}{5} \ln{\frac{M_U}{M_4}}
- \frac{24}{5} \ln{\frac{M_U}{M_5}}\\
\lambda_2&=&6 
- 50 \ln{\frac{M_U}{M_1}}
-40 \ln{\frac{M_U}{M_3}}
-30  \ln{\frac{M_U}{M_5}}
\\
\lambda_3&=& 5+ 21 
\ln{\frac{M_U}{M_R}} 
- 62\ln{\frac{M_U}{M_1}}
- 17 \ln{\frac{M_U}{M_2}}
- \frac{36}{5} 
\ln{\frac{M_U}{M_3}}
\\ \nonumber &&
- 12 \ln{\frac{M_U}{M_4}}
- 12 \ln{\frac{M_U}{M_5}},
\end{eqnarray}
where $M_U$ is the grand unification scale, $M_R$ is the mass of the
nine gauge bosons of SO(10) not contained in the standard model and
that do not lead to proton decay, $M_1$ is the mass of the scalars in
the (1,1,6) and (2,2,15) contained in the ${\bf 126}$, $M_2$ is the
mass of the scalars contained in the (1,3,10) contained in the ${\bf
126}$, $M_3$ is the mass of the scalars contained in the
(1,3,$\overline{10}$) contained in the ${\bf 126}$, $M_4$ is the mass
of the scalars in the (1,3,15) contained in the ${\bf 210}$ (they have
a negative impact on SO(10) unification and one thus expect
$M_U/M_4\sim 1$), $M_5$ is the mass of the scalars in the (3,1,15)
contained in the ${\bf 210}$ (they are beneficial for SO(10)
unification). Note that there is a typographical mistake in eq. 17 of
\cite{Lavoura:su}, a factor $-24/5 \ln(M_U/M_5)$ is missing in the
definition of $\lambda_1^S$.

For the proton lifetime estimate, we shall use:
\begin{eqnarray} \label{lifetime}
\tau_{\mbox{p $\to$ e$^+$ $\pi^0$}}&=& \frac{5}{8}
\left(\frac{\alpha_U^{\mbox{SU(5)}}}{\alpha_U^{\mbox{SU(10)}}}\right)^2
\\ \nonumber \times &&4.5 \times 10^{29} \left(\frac{M_U}{2.1 \times 10^{14}
\mbox{GeV}} \right)^4 \mbox{yr},
\end{eqnarray}
following Lee et al. \cite{Lee:1994vp} and assuming that
$\alpha_U^{\mbox{SU(5)}} \approx \alpha_U^{\mbox{SU(10)}}$.  For a
numerical estimate we use $\omega_1(M_Z)=1/0.016887$,
$\omega_2(M_Z)=1/0.03322$ and find that the set of parameters
$M_1=M_2=M_3=M_4=M_5=M_U$, $M_R=1/104 M_U$ lead to
$\alpha_s(M_Z)=0.120$, $M_U=4.3 \times 10^{15}$ GeV and
$\tau_{\mbox{proton}}=4.4 \times 10^{34}$ years. Another set of
parameters is e.g. $M_1=M_2=M_4=M_U$, $M_R=1/38 M_U$, $M_3=1/2.18 M_U$
and $M_5=1/2 M_U$ lead to $\alpha_s(M_Z)=0.120$, $M_U=2.95 \times
10^{15}$ GeV and $\tau_{\mbox{proton}}=1 \times 10^{34}$ years. It
seems difficult to push the proton lifetime above $10^{34}$
years. This is a prediction of that model with direct breaking to the
standard model at the grand unification scale. Besides neutrino
masses, the only new phenomenon is proton decay with a lifetime of
the order of $10^{34}$ years. Note that this is only one order of
magnitude above the present experimental limit for proton decay
\cite{Hagiwara:fs}. It is nevertheless possible to have a longer
proton lifetime if there is an intermediate scale
\cite{Mohapatra:1992dx,Lee:1994vp,Parida:1996td}. 

\subsection{Minimal models with two steps breaking of SO(10)}

Mohapatra and collaborators have studied these cases extensively. Four
different breaking schemes can be considered:
\begin{itemize}
\item[a)] SO(10) $\to$ $G_{224D}$=SU(2)$_L \times$ SU(2)$_R \times$ SU(4)$_C \times$ $D$

\item[b)] SO(10) $\to$  $G_{224D}$=SU(2)$_L \times$ SU(2)$_R \times$ SU(4)$_C$

\item[c)] SO(10) $\to$ $G_{2213D}$=SU(2)$_L \times$ SU(2)$_R \times$ U(1)$_{B-L} 
\times$ SU(3)$_C  \times$ $D$

\item[d)] SO(10) $\to$ $G_{2213}$=SU(2)$_L \times$ SU(2)$_R \times$
U(1)$_{B-L} \times$SU(3)$_C$,
\end{itemize}
assuming that the intermediate scale $M_I$ is where SO(10) is
broken. Case a) arises if the Higgs multiplet used is in the ${\bf
54}$. Cases b) and c) arise if a Higgs multiplet in the ${\bf 210}$ is
used. The nature of the intermediate gauge symmetry depends on the
details of the Higgs potential. Finally case d) arises if a {\bf 45}
and a {\bf 54} are used to break SO(10). In all cases a {\bf 126} and
a {\bf 10 }are needed to break the intermediate gauge symmetry to
U(1)$_{em}$ of QED. The predictions of each of these models for the
proton lifetime are \cite{Lee:1994vp}: a) $\tau_{p\to e^+ \pi^0}\sim
1.44 \times 10^{32}$ yr, b) $\tau_{p\to e^+ \pi^0}\sim 1.44 \times
10^{37.4}$ yr, c) $\tau_{p\to e^+ \pi^0}\sim 1.44 \times 10^{34.2}$
yr, d) $\tau_{p\to e^+ \pi^0}\sim 1.44 \times 10^{37.7}$ yr. The
uncertainties in these predictions have been discussed in
\cite{Lee:1994vp,Parida:1996td}. Despite these uncertainties, model a)
is probably excluded by direct searches for proton decay. A CP
violating phase in the CKM matrix compatible with present experiments
requires another multiplet e.g. a {\bf 120} \cite{Dutta:2004wv}. This
multiplet is assumed to be very heavy, i.e. its mass is of the order
of the grand unification scale, such that it does not contribute to
the running of the gauge coupling.

\subsection{Predictions of SO(10) grand unified models} 
\begin{itemize}
\item[a)] neutrino masses and oscillations are expected,
\item[b)] proton decay, the lifetime of the proton is around $10^{34}$
years if SO(10) is broken directly to the standard model at the GUT
scale or up to about $10^{38}$ years if there is an intermediate scale
at $10^{13}$ GeV,
\item[c)] one light Higgs boson will be observed at the LHC but no signal
for any new physics whatsoever.
\end{itemize}
These are the three firm predictions of a SO(10) grand unified theory
which is either directly broken at the grand unification scale to the
standard model or which is broken first to a subgroup and then at an
intermediate scale to the standard model.

It should be noted that although we are giving up to explain the
naturalness or fine-tuning problem, the gauge hierarchy problem can be
understood in the framework of a grand unified theory. A
renormalization group equation ``explains'' the hierarchy problem:
once the high scale value is fine-tuned, the low scale value of the
Higgs boson expectation value can be predicted.

Within string/M theory, where the landscape reasoning to solve the
fine-tuning problem makes sense, gauge couplings are expectation
values of moduli and can thus have a time dependence. As pointed out
in \cite{Calmet:2001nu}, if gauge couplings have a time dependence,
one might be able to obtain some information on the nature of the
grand unified theory. Although this prediction of these models is 
more speculative that the three predictions mentioned above, an
observation of a time dependence of the gauge couplings fulfilling the
relations derived in \cite{Calmet:2001nu} together with the
observation of only one Higgs boson at the LHC, would have to be
interpreted as a hint that the landscape scenario is a reasonable
explanation for the fine-tuning problem of the standard model.

\section{Conclusions}

If the world we live in is indeed fine-tuned, grand unification does
not require supersymmetry. Supersymmetry might still be necessary for
quantum gravity, but there is no good motivation to require that any
superpartner has a mass below the Planck scale. It might thus be a
hopeless task to detect any effect of supersymmetry. On the other
hand, proton decay is unavoidable and is a clear signature of a grand
unification. One of the predictions of SO(10) neutrino masses and
oscillations has already been observed. We have presented the minimal
models, one could imagine decoupling the different scales of the
models and introducing more scalar multiplets. It is interesting to
note that, once fine-tuning is allowed the only motivation for
supersymmetric unification is dark matter. Non-supersymmetric models
do not have ``natural'' candidates. But, without any further
experimental evidence, this remains a very weak motivation for low
energy supersymmetry or split supersymmetry. A interesting possibility
is that nature is indeed supersymmetric at the grand unification scale
and that there is a nearly exact chiral symmetry that protects the
supersymmetric dark matter candidate from developing a very large
mass, but that on the other hand the remaining fermionic superpartners
are very massive because their chiral symmetries are more strongly
broken. In that scenario, supersymmetry would only be required to
explain dark matter.

The LHC might just discover one single Higgs boson, this would be a
second evidence, after the cosmological constant, that the guidance
principle we had for model building was not the right one. This could
be explained by the anthropic principle, if we live in a landscape of
vacua or simply by the fact that renormalization is a physical
principle and that gauge and Yukawa couplings are just parameters of
the theory that have to be measured. As such their magnitudes, small
or large, do need not to be explained. The only physical question
remains to explain the splitting between the Planck scale and the weak
scale (gauge hierarchy problem), but this seems rather simple to
understand within the framework of a grand unified theory, as it would
be the consequence of gauge symmetry breaking and the running of the
parameters of the Higgs boson's potential from the grand unified scale
to the weak scale. Another interesting challenge is to understand how
to generate or trigger the weak phase transition. This is naturally
explained in supersymmetric models, but at the price of supersymmetry
breaking. As a conclusion, we want to emphasize that fine-tuning as a
guidance principle for searching for physics beyond the standard model
might not be the right one for different reasons and one should remain
very open minded when it comes to analyze the LHC data, as a complete
surprise is not that improbable.

\section{Acknowledgments}
The author is grateful to S.~Hsu, M.~Graesser, M.~B.~Wise and Z.~Xing
for useful discussions.

\end{document}